# A brief and transparent derivation of the Lorentz-Einstein transformations via thought experiments

Bernhard Rothenstein[1], Stefan Popescu [2] and George J. Spix [3]

1) Politehnica University of Timisoara, Physics Department, Timisoara, Romania
2) Siemens AG, Erlangen, Germany
3) BSEE Illinois Institute of Technology, USA

**Abstract.** *Starting with a thought experiment proposed by Kard[10], which derives the formula that accounts for the relativistic effect of length contraction, we present a "two line" transparent derivation of the Lorentz-Einstein transformations for the space-time coordinates of the same event. Our derivation make uses of Einstein's clock synchronization procedure.*

## 1. Introduction

Authors make a merit of the fact that they derive the basic formulas of special relativity without using the Lorentz-Einstein transformations (LET). Thought experiments play an important part in their derivations. Some of these approaches are devoted to the derivation of a given formula whereas others present a chain of derivations for the formulas of relativistic kinematics in a given succession. Two anthological papers1,2 present a "two line" derivation of the formula that accounts for the time dilation using as relativistic ingredients the invariant and finite light speed in free space $c$ and the invariance of distances measured perpendicular to the direction of relative motion of the inertial reference frame from where the same "light clock" is observed. Many textbooks present a more elaborated derivation of the time dilation formula using a light clock that consists of two parallel mirrors located at a given distance apart from each other and a light signal that bounces between when observing it from two inertial reference frames in relative motion.[3,4] The derivation of the time dilation formula is intimately related to the formula that accounts for the length contraction derived also without using the LET. The addition law for relativistic velocities can be derived without using the LET by using the formulas that account for time dilation and length contraction[5,6], from the invariance of $c$[7] or using the time dilation formula[8]. The scenarios following the derivation involve races between light signals and tardyons, presenting many turning points that make them hard to teach without mnemonic aids.

Formulas that account for relativistic effects are derived without using (LET) in the following succession: radar echo, time dilation, Doppler Effect, addition of relativistic velocities[9] or in the alternative succession: length contraction, Doppler Effect, addition of relativistic velocities and time dilation[10]. The striking fact is that some of the formulas derived without using the LET can be used to straightforwardly derive the LET without loosing in transparence. Authors[9,11] consider that the derivation of relativistic formulas using the LET is straightforward but rather formal and not very transparent from the point of view of physics. We consider that the LET are not only straightforward but also transparent when we have a correct representation about the physical meaning of the physical quantities involved by these transformations.

## 2. Length contraction without LET

The derivation of the fundamental formulas of special relativity teaches us that they are a direct consequence of the fact that the light propagates with finite and invariant speed throughout the empty space. The way in which Kard[10] derives the formula that accounts for the length contraction leads to a "two line" derivation of the LET for the space-time coordinates of the same event. It proceeds as follows: consider a pair of parallel rods in relative motion **1** and **2** (Figure 1). Measured by observers relative to whom the rods are in a state of rest their (proper) lengths are $L_{0,1}$ and $L_{0,2}$ respectively. Rod **1** moves with speed *V* relative to rod **2** whereas rod **2** moves with speed *–V* relative to rod **1**. The two rods are positioned along the overlapped axes OX(O'X') of their rest frames  K(XOY) for rod **2** and K'(X'O'Y') for rod **1**. The corresponding axes of the two frames are parallel to each other. The diagram shown in Figure 1 is a space-diagram reflecting the point of view of observers at rest in K. Therefore $L_{0,2}$ represents in true value the length of the stationary rod whereas $L_1(V)$ represents the length of the moving rod when measured by observers from K. At a time *t=0*  the left ends of both rods are located at the coinciding origins O and O' of the reference frames. At this very moment a source of light S located at O emits a light signal in the positive direction of the common axes. Consider the new situation occurring after a given time of motion when the right ends of both rods coincide in space. We have obviously

$$L_{0,2} = L_1(V) + \frac{V}{c}L_{0,2} \quad . \tag{1}$$

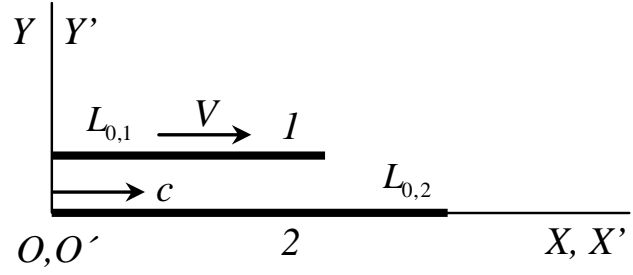

**Figure 1.** *Kard's scenario for deriving the length contraction formula without using the Lorentz-Einstein transformations*

The second term in the right side of (1) represents the distance traveled by the right end of rod **1** in order to arrive in front of the right end of rod **2**. Considering the same experiment from the reference frame K' we have obviously

$$L_{0,1} = L_2(V) - \frac{V}{c}L_{0,1} \tag{2}$$

The second term in the right side of (2) represents the distance traveled by the right end of rod **2** in order to arrive in front of the right end of rod **1**. From (1) we have

$$L_1(V) = L_{0,2}(1 - \frac{V}{c}) \tag{3}$$

whereas from (2) we obtain

$$L_2(V) = L_{0,1}(1 + \frac{V}{c}). \tag{4}$$

Combining (3) and (4) we obtain

$$\frac{L_1(V)}{L_{0,1}} \cdot \frac{L_2(V)}{L_{0,2}} = (1 - \frac{V^2}{c^2}). \tag{5}$$

Because

$$\frac{L_1(V)}{L_{0,1}} = \frac{L_2(V)}{L_{0,2}} \tag{6}$$

we conclude that in general the proper length of a rod $L_0$ and its length measured by observers relative to whom it moves with constant velocity $V$ are related by

$$\frac{L(V)}{L} = \sqrt{1 - \frac{V^2}{c^2}} \qquad (7)$$

a formula that accounts for the length contraction effect because as we may see $L(V) < L_0$.

## 3. Events and their space-time coordinates related by the Lorentz-Einstein transformations

The concept of event is fundamental in the special relativity theory being defined as "any physical occurrence that happens at a definite place in space and at a defined instant in time". The point where the event takes place is defined by its space coordinates with respect to a reference frame say K. The concept of rod length is intimately associated with the space coordinates that define the location of its ends. Consider a rod located along the OX axis. Its ends are $\mathbf{1}(x_1, 0)$ and $\mathbf{2}(x_2, 0)$ its length being defined by

$$L = (x_2 - x_1) \qquad (8)$$

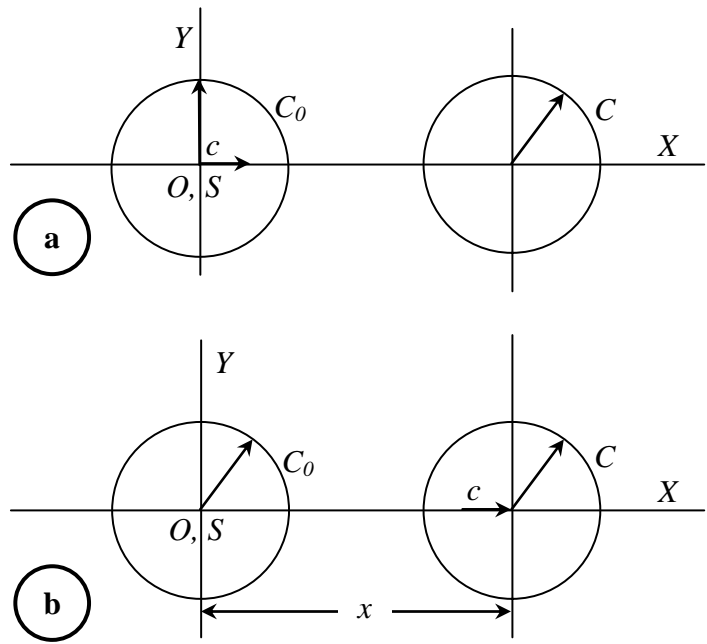

***Figure 2.*** *Synchronization of two distant clocks following Einstein's clock synchronization procedure*

Figure 2 shows a synchronization procedure of the clocks $C(0,0)$ and $C(x,0)$ in accordance with a clock synchronization procedure proposed by Einstein. Clock $C(0,0)$ ticks, clock $C(x,0)$ is stopped and set to display a time $t = \frac{x}{c}$ that is the time during which a light signal travels from the first clock to the second one. When $C_0(0,0)$ reads zero *t=0* a source of light S located in front of it emits a light signal in the positive direction of the OX axis. Arriving at the location of the second clock the light signal starts it and from this very moment the two clocks display the same running time being synchronized in accordance with the procedure proposed by Einstein. An identical procedure synchronizes the clocks $C_0'(0,0)$ and $C'(x,0)$ of the K frame.

Figure 3 shows the relative position of the reference frames K and K' when the right ends of rods **1** and **2** coincide in space as detected from K whereas Figure 4 shows the same situation as detected from the reference frame K'. In the case shown in Figure 3 clocks $C_0(0,0)$ and $C(x = L_{0,2},0)$ read both *t* as a result of the fact that the light signal emitted by the source S involved in the Kard[10] scenario synchronizes them via the light signal emitted at *t=0*. During the time interval (*t-0*) the origin O' has advanced with *V(t-0)*. The rod **1** is at rest in K' with its left end at O', its rest length being $L_1 = x' - 0$. Rod **2** has its left end at O its length being $L_{0,2} = x - 0$. Simply adding lengths measured in the same K frame we have

$$x = \sqrt{1 - \frac{V^2}{c^2}}x' + Vt \qquad (9)$$

wherefrom we obtain

$$x' = \frac{x - Vt}{\sqrt{1 - \dfrac{V^2}{c^2}}} = \gamma(x - c\beta t) \qquad (10)$$

with the usual shorthand notations $\gamma = (1 - \beta^2)^{-1/2}$ *and* $\beta = V/c$.

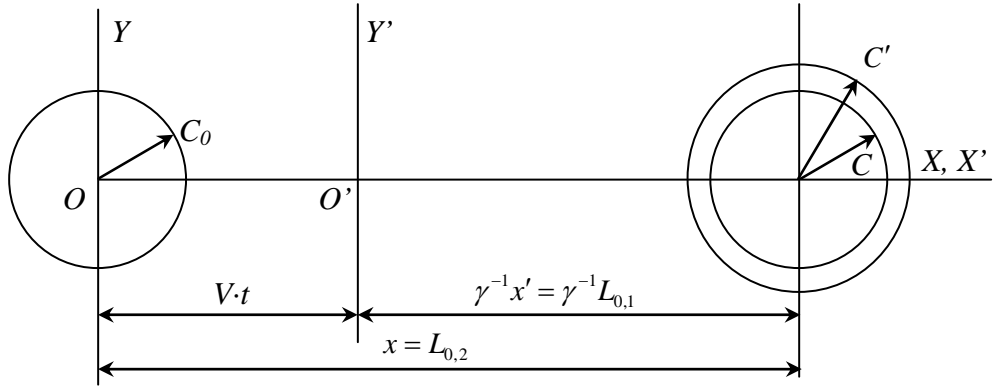

**Figure 3.** *Scenario for deriving the length contraction formula without using the Lorentz-Einstein transformation as detected from the K inertial reference frame.*

Figure 4 shows the same situation as detected from K'. Here the synchronized clocks $C_0'(0,0)$ and $C'(x',0)$ read both $t'$ when the origin O' is located at a distance $Vt'$ away from O'. The length of rod **2** measured by observers from K is $\gamma^{-1}L_{0,2} = \gamma^{-1}x$. Adding only lengths measured in K' we obtain

$$\gamma^{-1}x = x' + \beta ct' \qquad (11)$$

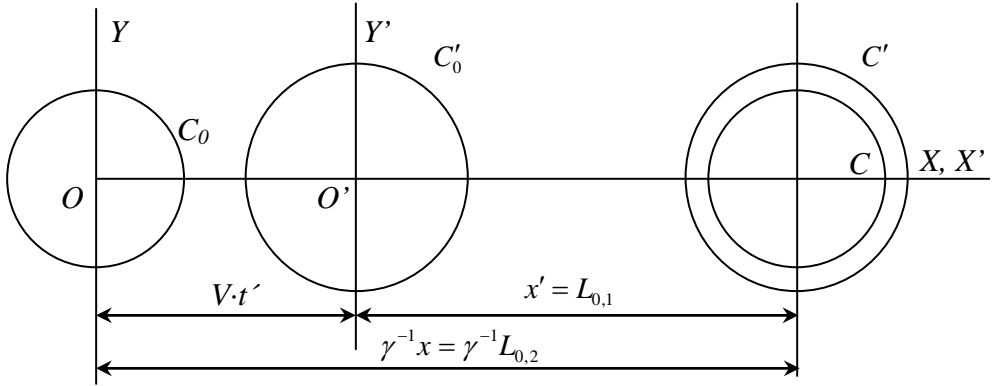

***Figure 4.*** *Scenario for deriving the length contraction formula without using the Lorentz-Einstein transformations as detected from the K' inertial reference frame.*

wherefrom we obtain

$$x = \gamma(x' + c\beta t') . \qquad (12)$$

Combining (10) and (12) we obtain

$$t = \gamma(t' + \frac{\beta}{c}x') \qquad (13)$$

and

$$t' = \gamma(t - \frac{\beta}{c}x) . \qquad (14)$$

Equation (1) combined with the formula that accounts for length contraction leads directly to the LET for the time coordinates of the same event. Expressing (1) as a function of space coordinates it becomes

$$x = x'\sqrt{1 - \beta^2} + \beta x . \qquad (15)$$

Divided by the invariant $c$ and taking into account that the synchronization of the involved clocks took place *à la* Einstein ($t = \dfrac{x}{c}; t' = \dfrac{x'}{c}$) (15) leads to (14). Handled in the same way, (2) leads to (13).

The space coincidence of the right ends of the rods generates an event. Detected from K this event is characterized by its space coordinates (x,y=0) and by its time coordinate $t$ that equates the reading of clock $C(x.0)$ located at the point where the event takes place when it takes place. The notation $E(x,0,t)$ stands for the event in K. Detected from K' same the event is defined by $E'(x',0,t')$, the space coordinates $(x',0)$ defining the point where it takes place, whereas $t'$ equates the reading of clock $C'(x',0)$ located at this point when the event takes place. Because events the $E$ and $E'$ take place at the same point in space when the clocks $C$ and $C'$ read $t$ and $t'$ respectively, the relativists say that they represent the same event.

Equations (12), (13), (10) and (14) represent the celebrated LET for the space-time coordinates of the same event detected from two inertial reference frames being in relative motion with speed $V$.

Knowing the physics behind the Lorentz-Einstein transformations and having a correct representation about the meaning of the physical quantities they relate, we can use them for deriving formulas that account for the fundamental relativistic effects without obscuring the physics behind them.

### 4. Conclusions

Starting with a thought experiment proposed by Kard[10], which derives the formula that accounts for the relativistic effect of length contraction without using the LET, we have included the distant clock synchronization procedure proposed by Einstein thereby obtaining a simple and transparent derivation for the LET.